\documentclass[a4paper,10pt]{article}
\usepackage[utf8]{inputenc}
\usepackage{amsmath}
\usepackage{amsfonts}
\usepackage{float}
\usepackage{graphicx}

\newcommand{\no}[1]{\lVert #1 \rVert}

\def\spvec#1{\left(\vcenter{\halign{\hfil$##$\hfil\cr \spvecA#1;;}}\right)}
\def\spvecA#1;{\if;#1;\else #1\cr \expandafter \spvecA \fi}

\title{Convergent series for polynomial lattice models with complex actions}
\author{
Vasily Sazonov\\
Laboratoire de Physique Th\'eorique, CNRS UMR 8627,\\ 
Universit\'e Paris XI,  F-91405 Orsay Cedex, France}

\begin{document}

\maketitle

\begin{abstract}
Lattice models with complex actions are 
important for the understanding of matter at finite densities, but
not accessible by the standard Monte Carlo techniques due to the sign problem.
Here we derive a new approach for avoiding the complex action/sign problem, by
extending the method of convergent series with a non-Gaussian
initial approximation. The main features of the new series are demonstrated on the example 
of the two dimensional oscillating integral.
\end{abstract}

% \begin{abstract}
% Recently, the convergent series employing the non-Gaussian initial approximation
% were constructed for polynomial lattice models with real actions and were shown to be 
% an effective computational tool. Here we extend the convergent series derivation 
% to a class of polynomially interacting models with complex actions and 
% demonstrate main features of the series on a simple example.
% \end{abstract}

\section{Introduction}
One of the most traditional approaches to computations of the path integral or of its discretized lattice version
is the standard perturbation theory (SPT), a perturbative expansion around the Gaussian initial approximation.
Significant progress has been made in calculations of the standard quantum field
perturbation theory in the last few years. There were developed numerical and analytical methods for calculations of multi-loop
Feynman diagrams \cite{AKNS2013, Panzer2015, Golz2015}, which led to the evaluation of approximations 
with a record-breaking number of loops \cite{Adzhemyan2013, Batkovich2016, Kompaniets2017, Oliver}.

However, the series of SPT are asymptotic \cite{Dyson, Lipatov} 
and can be used only for small expansion parameters (coupling constants). 
The reason is in the incorrect interchange of the summation and
integration during the construction of the perturbative expansion. 
From a physical point of view, the divergence of the perturbative series is related
to large fluctuations of fields.
Nevertheless, one can avoid the problem of the interchange by choosing a proper non-Gaussian initial approximation.
\footnote{For the alternative methods see \cite{MeuricePRL, Meurice2004, Belokurov1, Belokurov2, ZeroDimLVE, HowToResum}.}

The convergent series (CS) based on the substitution of 
the Gaussian initial approximation by a certain interacting theory was initially 
proposed for the quantum anharmonic oscillator and scalar field theories \cite{Halliday, Ushveridze1983, Shaverdyan1983, UshveridzeSuper}
and later developed in \cite{Turbiner, SISSAKIAN1992, VPTSolovtsov, Nalimov}. 
The main feature of the CS method is that the employed non-Gaussian initial approximation 
can be reduced by a set of transformations to Gaussian integrals. Consequently, each order of CS is expressed in terms of a linear 
combination of a finite amount of orders of the standard perturbation theory.
However, the derivation of the convergent series for quantum field theories is based on the assumption of the
applicability of the dimensional regularization \cite{Leibbrandt}
to handle the limit of the infinite number of degrees of freedom.
Due to this fact a rigorous mathematical proof of the series convergence is still missing.

Recently, the convergent series 
similar to \cite{Shaverdyan1983, UshveridzeSuper} was derived for the real action models defined 
on finite lattices  \cite{IvanovProc, CSLatticePhi4}. There the restriction to a finite amount
of degrees of freedom provided the conditions, enough to carry out the construction mathematically rigorously and
to prove the possibility to express coefficients of the convergent series for any model on the finite lattice
with the real action and an even degree polynomial interaction
as linear combinations of SPT-terms. 
It was also shown in \cite{CSLatticePhi4} that the convergent series has a certain variational invariance,
which allows one to improve its convergence significantly. 

In the current paper we extend the non-perturbative 
convergent series method to a class of lattice models with complex actions.
For such models the standard Monte Carlo approach is not applicable because of the
lack of the Boltzmann weight positivity (complex action problem or sign problem). 
The sign problem is typical for theories describing matter at non-zero densities
and for models with a vacuum term. 
% For instance, it hinders our understanding of the QCD phase diagram and 
% high-temperature superconductivity. 
Different approaches are studied nowadays in attempts to overcome it 
\cite{ApproachesSP, Borsanyi2015, Sexty2014, Gattringer2014, Aarts2013}. Nevertheless, all of them have certain limitations
and the development of new methods for models with complex actions is still particularly important.
Hereafter, we consider sign problems related to the presence of finite chemical potential and, 
therefore, assume that the complex contributions are contained only in the Gaussian part of the action.
We construct the convergent series with a non-Gaussian initial approximation for lattice models with polynomial 
interactions and complex quadratic part of the action and prove the existence of the variational invariance, 
highly important for the applicability of the method.
We demonstrate the work of the method on the example of the oscillating two dimensional integral.

\section{Convergent series}
Consider a model defined on the $d$-dimensional cubic lattice, with $N^d$ lattice cites,
by the action
\begin{equation}
  S[\phi] = \sum_{x,y}\Big(\phi^T_x K_{xy} \phi_{y}
  + g \sum_x (\phi^T_x\phi_x)^{p}\Big)\,.
\label{S}
\end{equation}
Here $\phi_x$ is a real vector of the dimension $\tilde d$ (the vector indices are included in $x$) 
$\phi^T_x$ is a transposition of $\phi_x$, $x$ and $y$ run over all lattice sites and vector indices of fields,
$g > 0$ is a coupling constant, $p > 1$ characterizes the degree of the interaction, $K_{xy}$ is some complex matrix
with a number of eigenmodes given by $V = N^{d\tilde d}$, the latter quantity corresponds to .
Obviously, a wide class of models is covered by the action (\ref{S}), for instance, the models of the scalar 
charged particles can be represented by (\ref{S}) with $\tilde d = 2$.

The partition function corresponding to the action (\ref{S}) is given by
\begin{equation}
  Z =  \int [d\phi]\,e^{-S[\phi]}\,,~~~\int [d\phi] = \prod_x \Big(\int [d\phi_x]\Big)\,.
\label{Z}
\end{equation}
% and the free energy is
% \begin{equation}
%   F = -\frac{1}{V} \log(Z)\,.
% \label{F}
% \end{equation}
To construct the convergent expansion for (\ref{Z}), 
we choose a non-Gaussian initial approximation
\begin{equation}
  N[\phi] = \alpha \no{\phi}^2 + \sigma \no{\phi}^{2 p}\,,
\end{equation}
where $\no{\phi} = \sqrt{\phi_x^* M_{xy} \phi_y}$ is a norm in $\mathbf{R}^{N^{d\tilde d}}$ determined by
a hermitian matrix with positive eigenvalues, $M_{xy}$.
The parameter $\alpha \in \mathbf{C}$, $Re\{\alpha\} > 0$ and $\sigma > 0$ is 
fixed by the inequality  
\begin{equation}
\sigma \no{\phi}^{2 p} \geq g \sum_x (\phi^T_x\phi_x)^{p}\,,
\label{NgS}
\end{equation}
which always holds for $\sigma$ big enough. 
Then, we expand the partition function (\ref{Z}) into a series as
\begin{eqnarray}
  Z = \sum_{n} \frac{1}{n!} \int [d\phi]\,e^{-N[\phi]} (N[\phi] - S[\phi])^n\,.
\label{Z1}
\end{eqnarray}
When the inequality (\ref{NgS}) is valid, the series (\ref{Z1}) is absolutely convergent 
\begin{eqnarray}
\nonumber
  \sum_{n} \Big| \frac{1}{n!} \int [d\phi]\,e^{-N[\phi]} (N[\phi] - S[\phi])^n\Big| 
  \leq
\\ 
\nonumber
  \sum_{n} \frac{1}{n!} \int [d\phi]\,e^{-Re\{N[\phi]\}}  \big|N[\phi] - S[\phi]\big|^n 
  \leq 
\\
\nonumber
  \sum_{n} \frac{1}{n!} \int [d\phi]\,e^{-Re\{N[\phi]\}} (\big|\alpha \no{\phi}^2 - \sum_{x,y}\phi^T_x K_{xy} \phi_{y}\big|
  + \sigma \no{\phi}^{2 p} - g \sum_x (\phi^T_x\phi_x)^{p})^n\\
% \nonumber
  = \int [d\phi]\,e^{-Re\{\alpha\} \no{\phi}^2 + |\alpha \no{\phi}^2 - \sum_{x,y}\phi^T_x K_{xy} \phi_{y}| - g \sum_x (\phi^T_x\phi_x)^{p}} < \infty\,.
\label{est1}
\end{eqnarray}

To evaluate terms of the series (\ref{Z1}), we change $\no{\phi}$ to a positive one dimensional
variable $t$ by introducing auxiliary integration with the delta function
\begin{eqnarray}
  Z = \sum_{n} \frac{1}{n!} \int [d\phi]\,\int_0^\infty dt\, \delta(t - \no{\phi}) e^{-\alpha t^2 - \sigma t^{2 p}} 
  \Big(P_2[\phi] + \sigma t^{2p} - g \sum_x (\phi^T_x\phi_x)^{p}\Big)^n\,, 
\label{Z2}
\end{eqnarray}
where $P_2[\phi]$ is the quadratic part of the perturbation,
\begin{equation}
  P_2[\phi] = \alpha \no{\phi}^2 - \sum_{x,y}\phi^T_x K_{xy} \phi_{y}\,.
\end{equation}
Then, rescaling fields as $\phi_x^{new} = t \phi_x^{old}$ and denoting $\phi_x^{new}$ as $\phi_x$,
we obtain
\begin{eqnarray}
\nonumber
  Z = \sum_{n} \frac{1}{n!} \int [d\phi]\,\int_0^\infty dt\, \delta(t - t\no{\phi})
  \\
\nonumber
  \cdot e^{-\alpha t^2 - \sigma t^{2 p}} 
  \Big(t^2 P_2[\phi] + t^{2p}(\sigma - g \sum_x (\phi^T_x\phi_x)^{p})\Big)^n = \\
\nonumber
  \sum_{n} \sum_{k = 0}^n \frac{1}{n!} \spvec{n;k} 
  I_{2p}[2 k + 2p (n-k), \alpha, \sigma, V]\\
\nonumber
  \cdot \int [d\phi]\,\delta(1 - \no{\phi}) 
  (P_2[\phi])^{k} (\sigma - g \sum_x (\phi^T_x\phi_x)^{p})^{n-k} = \\
\nonumber
  \sum_{n} \sum_{k = 0}^n \sum_{l = 0}^k \frac{\sigma^{n-k-l}}{n!} \spvec{n;k} \spvec{n-k;l}
  I_{2p}[2 k + 2p (n - k), \alpha, \sigma, V] \\
  \cdot 
  \int [d\phi]\,\delta(1 - \no{\phi}) 
  (P_2[\phi])^{k} \big(-g\sum_x (\phi^T_x\phi_x)^{p}\big)^l\,,
\label{Z3}
\end{eqnarray}
where $I_{2p}[r, \alpha, \sigma, V]$ is the integral over the variable $t$, given by
\begin{equation}
  I_{2p}[r, \alpha, \sigma, V] = \int_0^\infty dt\, t^{V - 1 + r}\, e^{-\alpha t^2 - \sigma t^{2 p}}\,.
\label{I2p}
\end{equation}
The expression (\ref{Z3}) can be reduced to standard Gaussian integrals. Employing
the identity
\begin{eqnarray}
\nonumber
  \int [d\phi]\, \delta(1 - \no{\phi})\, \phi_{x_1}...\phi_{x_r}\,\phi^T_{y_1}...\phi^T_{y_q} = \\
  \frac{1}{I_2[r+q, \alpha, V]} \int [d\phi]\, \exp\{-\alpha \no{\phi}^2\}\, \phi_{x_1}...\phi_{x_r} \phi^T_{y_1}...\phi^T_{y_q}\,,
\label{idnt}
\end{eqnarray}
where 
\begin{equation}
  I_2[r+q, \alpha, V] = \int_0^\infty dt\, t^{V - 1 + r + q}\, e^{-\alpha t^2}\,,
\end{equation}
one  obtains
\begin{eqnarray}
\nonumber
  Z = \sum_{n} \sum_{k = 0}^n \sum_{l = 0}^k \frac{\sigma^{n-k-l}}{n!} \spvec{n;k} \spvec{n-k;l}
  \frac{I_{2p}[2 k + 2p (n - k), \alpha, \sigma, V]}{I_2[2 k + 2p l, \alpha, V]} \\
  \cdot \int [d\phi]\,e^{-\alpha \no{\phi}^2} 
  (P_2[\phi])^{k} \big(-g\sum_x (\phi^T_x\phi_x)^{p}\big)^l\,.
\label{Z4}
\end{eqnarray}
The explicit dependence on the number of degrees of freedom $V$ in (\ref{Z3}), (\ref{idnt}) and (\ref{Z4}) 
appears from the Jacobian of the rescaling of fields. As it was numerically demonstrated for the real lattice $\phi^4$-model
in \cite{IvanovProc} and \cite{CSLatticePhi4}, this explicit dependence causes the dramatical slowing down of the convergence
at large lattice volumes.
However, in \cite{CSLatticePhi4} it was proved that for the real $\phi^4$-model the formula,
analogous to (\ref{Z4}), is invariant under the substitution of $V$ by a variational parameter $\tau \in [1; \infty)$. Here
we prove that a similar invariance holds for the complex case.
Let $m = n - k - l$, then, (\ref{Z4}), with $V$ changed by $\tau$, becomes
\begin{eqnarray}
\nonumber
  Z = \sum_{m,k,l = 0}^\infty \frac{\sigma^{m}}{(m+k+l)!} \spvec{m+k+l;k} \spvec{m+l;l}
  \frac{I_{2p}[2 k + 2p (m+l), \alpha, \sigma, \tau]}{I_2[2 k + 2p l, \alpha, \tau]} \\
\nonumber
  \cdot \int [d\phi]\,e^{-\alpha \no{\phi}^2} 
  (P_2[\phi])^{k} \big(-g\sum_x (\phi^T_x\phi_x)^{p}\big)^l\\
\nonumber
\simeq \sum_{k,l = 0}^\infty \frac{1}{k!\,l!}
  \frac{\int_0^\infty dt\, \sum_{m=0}^\infty \frac{(\sigma t^{2 p})^{m}}{m!} 
  t^{\tau - 1 + 2 (k + p l)} \, e^{-\alpha t^2 - \sigma t^{2 p}}}{I_2[2 k + 2p l, \alpha, \tau]} \\
\nonumber
  \cdot \int [d\phi]\,e^{-\alpha \no{\phi}^2} 
  (P_2[\phi])^{k} \big(-g\sum_x (\phi^T_x\phi_x)^{p}\big)^l\\
= \sum_{k,l = 0}^\infty \frac{1}{k!\,l!}
  \int [d\phi]\,e^{-\alpha \no{\phi}^2} (P_2[\phi])^{k} \big(-g\sum_x (\phi^T_x\phi_x)^{p}\big)^l\,.
\label{Z5}
\end{eqnarray}
The sign $\simeq$ in (\ref{Z5}) stands to indicate, that two sides of the equation are 
only perturbatively equivalent, i.e. have the same standard perturbative expansions,
but may differ by some non-analytic contributions like $e^{-\frac{1}{g}}$. To prove the invariance of (\ref{Z4})
under the substitution of $V$ by $\tau$ non-perturbatively, we may represent the expansion (\ref{Z1}) as a result of consequent expansions
of the quadratic and interacting parts of the exponent $e^{N[\phi] - S[\phi]}$. Performing the first of them, we obtain an
absolutely convergent series
\begin{equation}
  Z = \sum_{n} \frac{Z^{(n)}}{n!} = 
  \sum_{n} \frac{1}{n!} 
  \int [d\phi]\,,(P_2[\phi])^n e^{-\alpha \no{\phi}^2 - \sigma \no{\phi}^{2 p} + (\sigma \no{\phi}^{2 p} - g\sum_x (\phi_x \phi_x^T)^p)}\,.
\label{P2E}
\end{equation}
Each term $Z^{(n)}$ can be considered separately and interpreted as an average of the operator $(P_2[\phi])^n$. 
Then,
\begin{equation}
  Z^{(n)} = \sum_{k} \frac{1}{k!} 
  \int [d\phi]\,,(P_2[\phi])^n e^{-\alpha \no{\phi}^2 - \sigma \no{\phi}^{2 p}}
  (\sigma \no{\phi}^{2 p} - g\sum_x (\phi_x \phi_x^T)^p)^k
\label{P2E2}
\end{equation}
and all other steps of the CS construction can be done.
The non-perturbative independence on $\tau$ of each $Z^{(n)}$ can be proved 
exactly, as in \cite{CSLatticePhi4} for the lattice $\phi^4$-model
with the real action.

Concluding this section, we would like to note that, the expansion (\ref{P2E}) is
similar to the Taylor expansion in chemical potential (TE) \cite{Gavai2004}.
It is well known, that TE for the connected correlation functions necessarily brakes down at
the chemical potential corresponding to the first singularity of connected functions. 
Such singularities in connected
functions are related to zeros of the partition function 
(generation functional of full correlation functions) and for the models with complex
actions can exist even at finite lattice volumes. 
However, the expansion (\ref{P2E}) is different from TE, due to the flexibility of the CS method, 
providing a wide range of possible choices for the initial approximation. 
The initial approximation of CS can even contain
oscillating factors regulated by the value of $Im\{\alpha\}$. The latter
fact significantly increases the potential applicability of CS.

\section{Example}
We demonstrate the application of the convergent series considering the two dimensional
oscillating integral
\begin{equation}
  Z_2 = \int dx \int dy\, e^{-x^2 - y^2 + i\mu\, x\, y - g (x^4 + y^4)}\,.
\end{equation}
To calculate the integral we choose the initial approximation as
\begin{equation}
  N = (x^2 + y^2) + g (x^2 + y^2)^2\,,
\end{equation}
which corresponds to $\alpha = 1$ and $\sigma = g$ (and is not the only one possibility).

In Fig. 1 we present a comparison of the numerically integrated $Z_2$ with the results obtained
by the convergent series with $10$ terms (as well as for plots from Fig. 2 and Fig. 3), 
depending on $g$ at several values of $\mu$. Numerical and CS computations in Fig. 1 are
in the excellent agreement. In Fig. 2 in a region of small $g$ a deviation 
of the CS results from the correct answer is observed. The reason is that at small $g$ the expansion
(\ref{Z4}) is more sensitive to large $\mu$ and one needs more terms of the series 
(or alternatively, one has to choose a more appropriate initial approximation) 
to reproduce the correct result.
The dependences of $Z_2$ on $\mu$ at $g = 0.1$ and $g = 1$, presented in Fig.~3, support the same observations.
In Fig. 4 we study the non-perturbative independence
on the variational parameter $\tau$. Three lines corresponding to $5$, $10$ and $20$ computed CS terms show,
that the sum of the series (\ref{Z4}), in which $V$ is substituted by $\tau$, tends to be a constant
with respect to the variations of $\tau$ with increasing number of terms, taken into account.
The evolution of the value of $Z_2$ depending on number of CS terms at different $\tau$ is shown in Fig. 5.

\begin{figure}[H]
\begin{center}
\includegraphics[height=6cm, width=8cm]{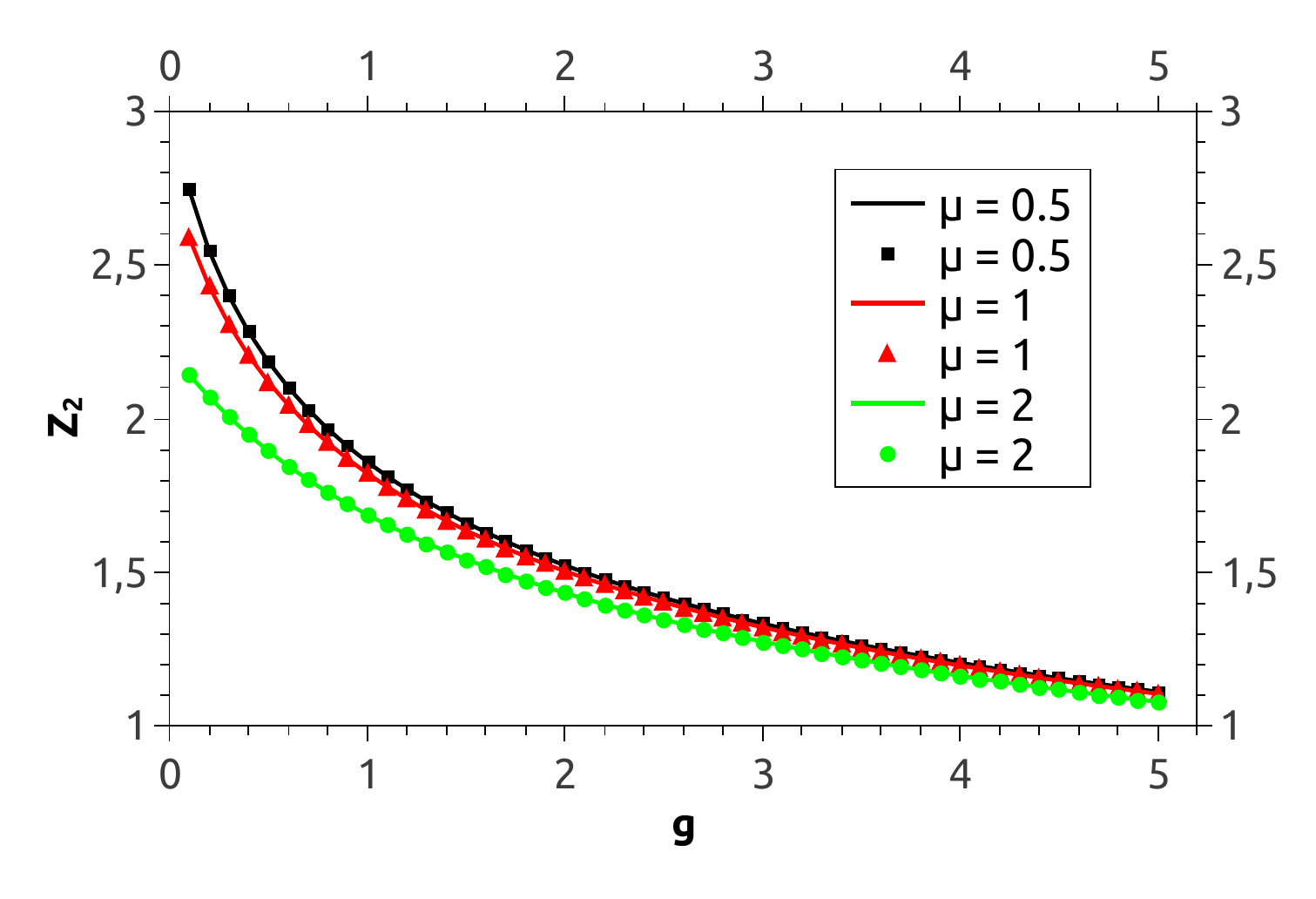}
\caption{The integral $Z_2$ depending on $g$ at $\mu = 0.5$ (black), $\mu = 1$ (red) and 
$\mu = 2$ (green). Solid lines correspond to the numerical integration
and bars, triangles and filled circles to results of CS.}
\end{center}
\label{fig:F1}
\end{figure}

\begin{figure}
\begin{center}
\includegraphics[height=6cm, width=8cm]{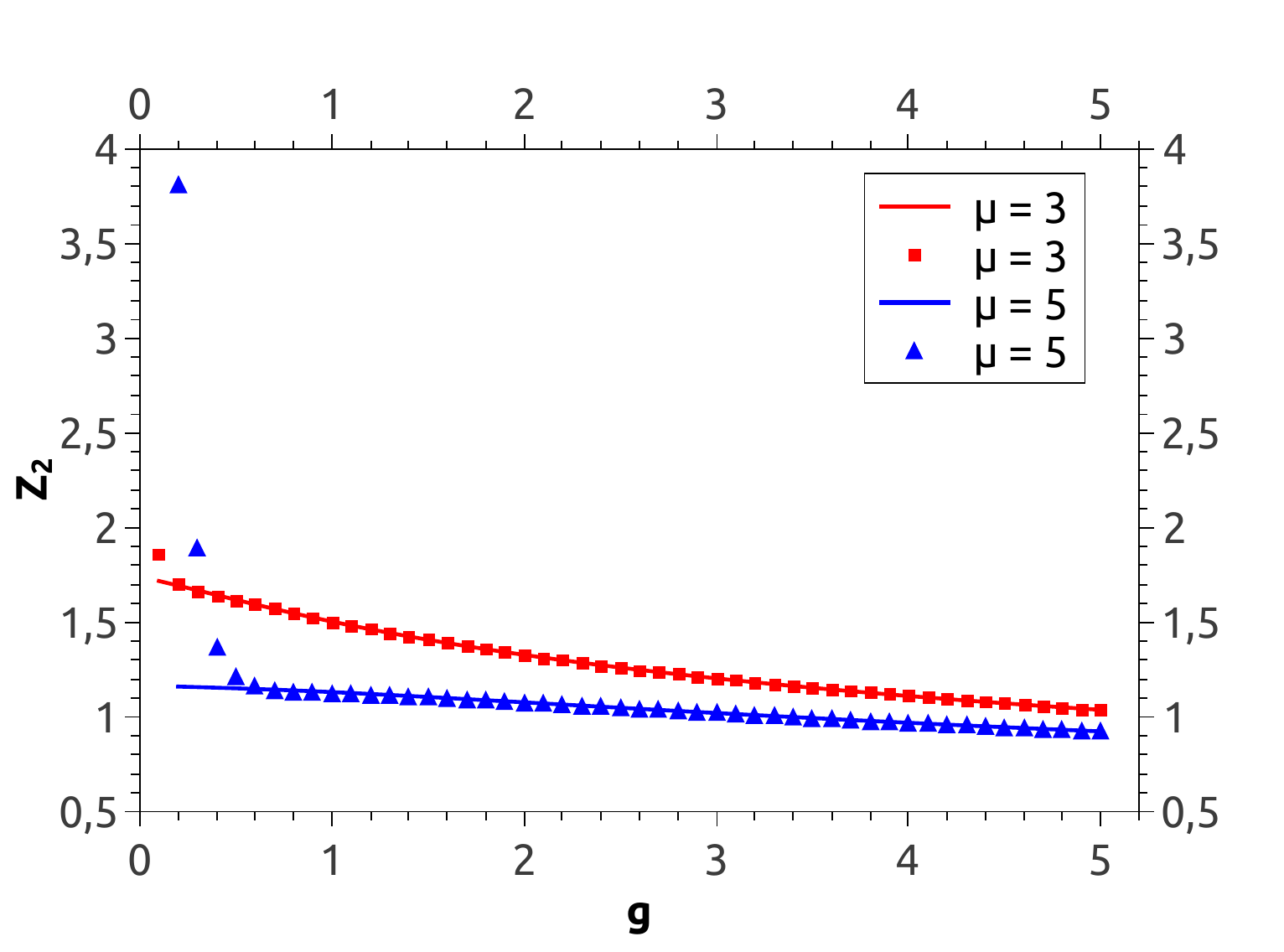}
\caption{The integral $Z_2$ depending on $g$ at $\mu = 3$ (red) and $\mu = 5$ (blue). 
Solid lines correspond to the numerical integration,
bars and triangles to results of CS.}
\end{center}
\label{F2}
\end{figure}

\begin{figure}
\begin{center}
\includegraphics[height=6cm, width=8cm]{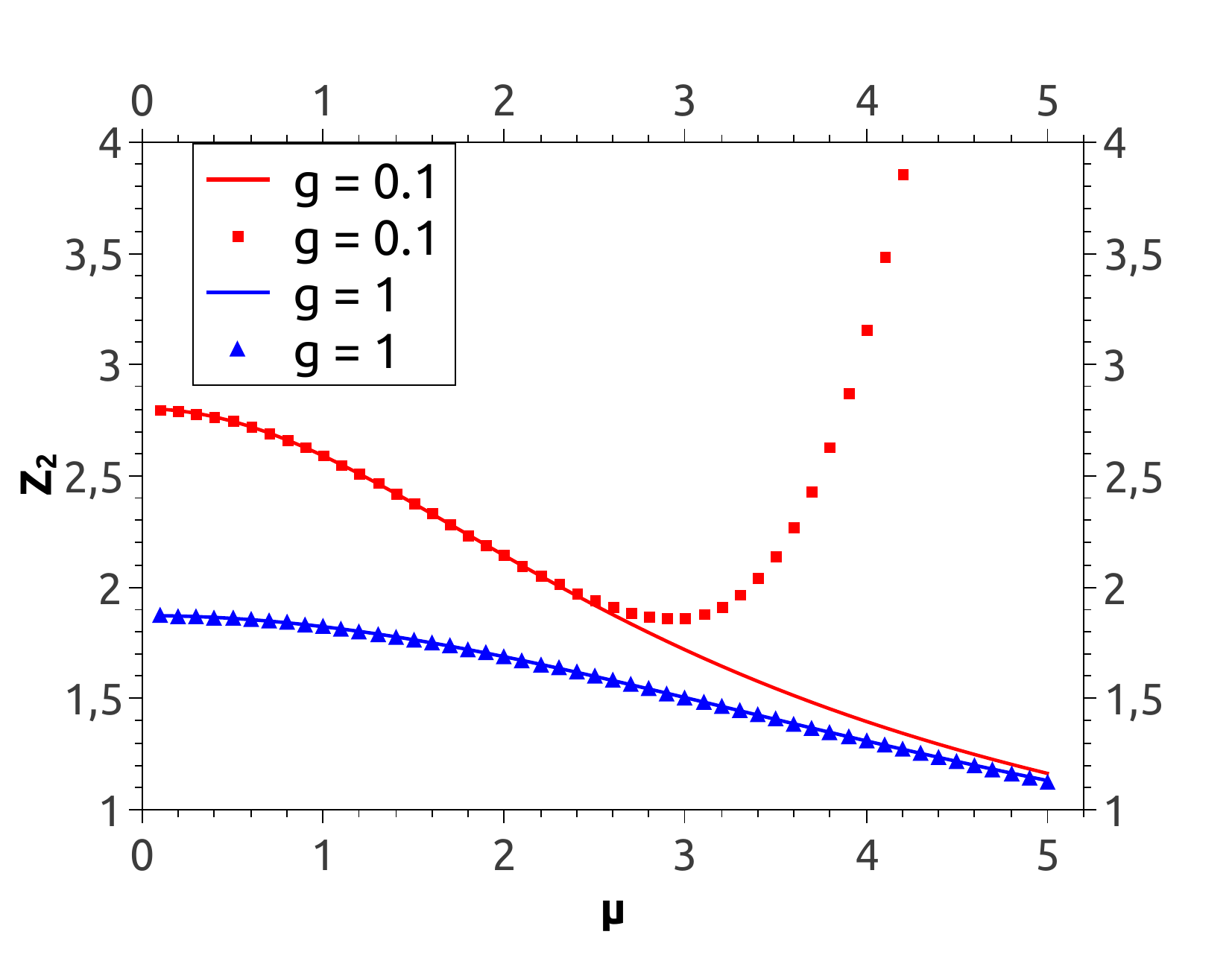}
\caption{The integral $Z_2$ depending on $\mu$ at $g = 0.1$ (red) and $g = 1$ (blue). 
Solid lines correspond to the numerical integration,
bars and triangles to results of CS.}
\end{center}
\label{F3}
\end{figure}

\begin{figure}
\begin{center}
\includegraphics[height=6cm, width=8cm]{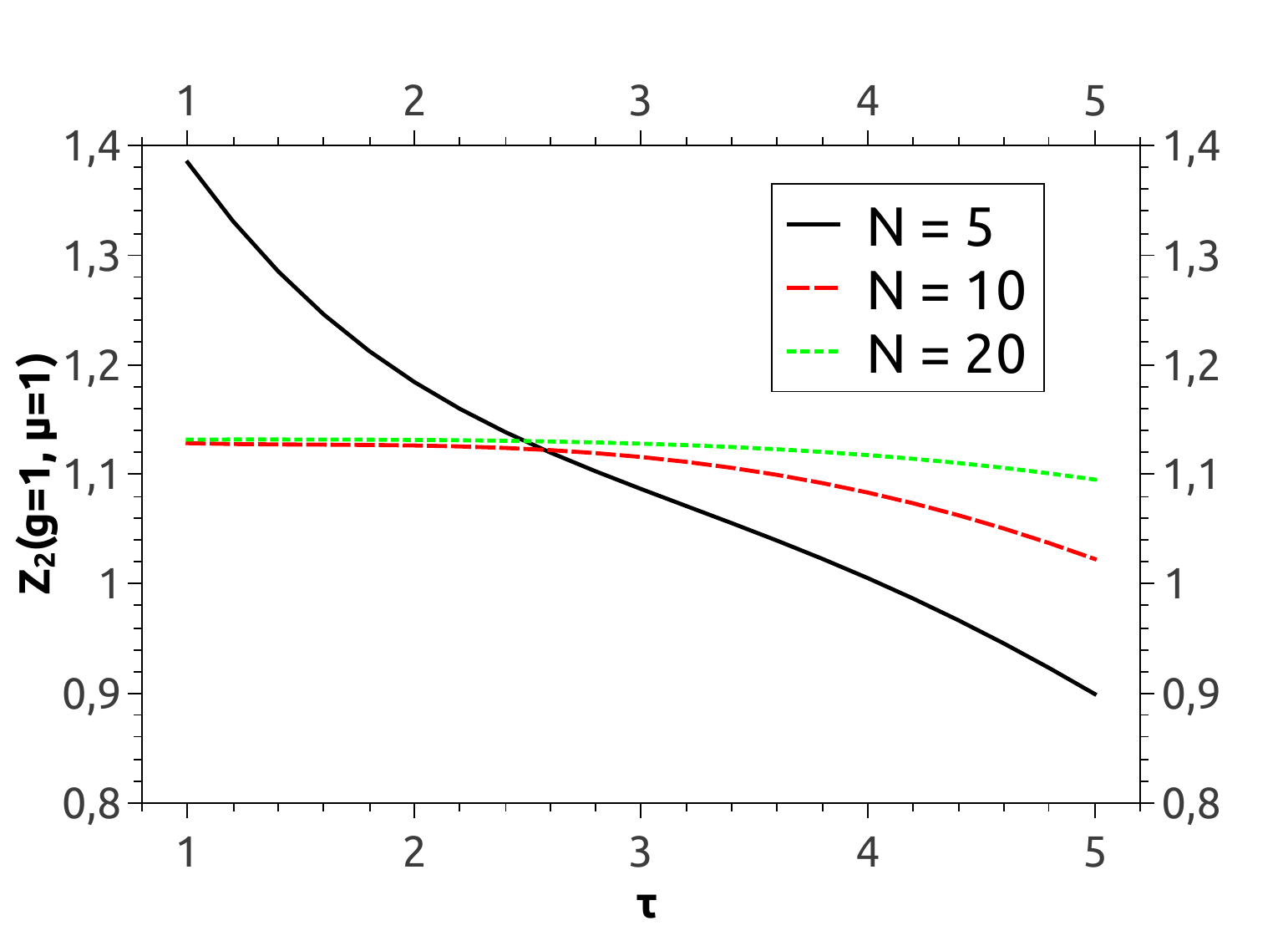}
\caption{The integral $Z_2$ at $g = 1$ and $\mu = 1$ depending on the variational parameter $\tau$
evaluated by $n = 5$ (solid black line) $n = 10$ (red dashed line) and $n = 20$ (green dotted line) terms of CS.}
\end{center}
\label{F4}
\end{figure}

\begin{figure}
\begin{center}
\includegraphics[height=6cm, width=8cm]{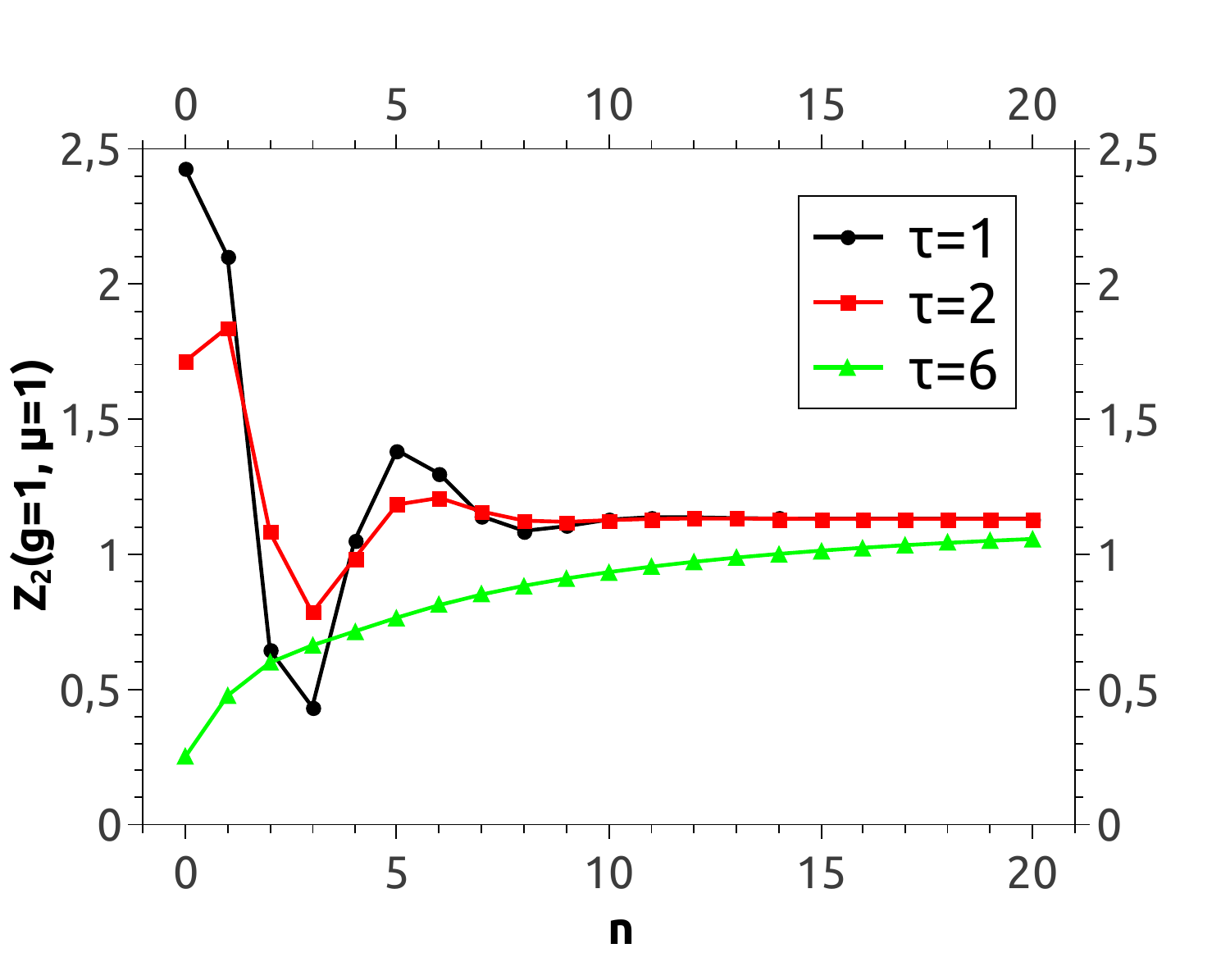}
\caption{The integral $Z_2$ at $g = 1$ and $\mu = 1$ depending on the number of computed terms of CS 
at variational parameters $\tau = 1$ (black line with filled circles), $\tau = 2$ (red line with bars) 
and $\tau = 6$ (green line with triangles).
}
\end{center}
\label{F5}
\end{figure}

\newpage
\section{Summary}
We have constructed the convergent series for lattice models with polynomial interaction and complex actions.
Within the CS framework the observables are expressed as sums of Feynman diagrams,
what reduces the computations to standard techniques.
The final expression for the series provides a possibility to introduce a variational parameter,
allowing one to exclude an explicit dependence on the number of degrees
of freedom and to improve the convergence of the method. 

The main objects of computations and measurements on the experiment are the connected correlation
functions. In the current paper we presented the construction of CS for the partition function.
The series for corresponding full correlation functions are derived exactly in the same way
and CS for the connected functions can be obtained by keeping only connected Feynman diagrams in the sum of the series.

The numerical results obtained for the simple example together with a variational invariance, excluding explicit
dependence on the number of degrees of freedom, open a new potential way to avoid the sign problem.

\subsection*{Acknowledgments}
The work was supported by the Austrian Science Fund (FWF) trough the Erwin Schr\"odinger fellowship J-3981.

\newpage
\label{Bibliography}
% \bibliographystyle{unsrt}
% \bibliography{bibliography}

\end{document}